\documentclass[twocolumn,showpacs,amsmath,amssymb]{revtex4-2}
\usepackage[T1]{fontenc}
\usepackage{hyperref}
\input{epsf}

\usepackage{graphicx}
\usepackage{epstopdf}
\usepackage{array}
\usepackage{longtable}
\usepackage{rotating,booktabs}
\usepackage{booktabs,threeparttable}
\usepackage{bm}
\usepackage{float}
\usepackage{amsmath}
\usepackage{gensymb}
\usepackage{multirow}

\usepackage{color}

\begin{document}

\title{The Land$\acute{e}$ $g$ factors for the $6S_{1/2}$ , $5D_{3/2,5/2}$ states of  Ba$^{+}$ ions }
\author{ Bing-Bing Li$^{1}$}
\author{Jun Jiang$^{1}$}
\email {phyjiang@yeah.net}
\author{Lei Wu$^{1}$}
\author{Deng-Hong Zhang$^{1,2}$}
\author{Chen-Zhong Dong$^{1}$}

\affiliation{$^{1}$Key Laboratory of Atomic and Molecular
Physics and Functional Materials of Gansu Province,
College of Physics and Electronic Engineering,
Northwest Normal University, Lanzhou 730070, P. R. China
}

\affiliation{$^{2}$College of Electronic Information and Electronic Engineering, Tianshui Normal University, Lanzhou 741000, P. R. China
}


\begin{abstract}
 The Land$\acute{e}$ $g$ factors of Ba$^+$ are very important in high-precision measurement physics. The wave functions, energy levels, and Land$\acute{e}$ $g$ factors for the $6s$ $^{2}S_{1/2}$ and $5d$ $^{2}D_{3/2,5/2}$ states of Ba$^{+}$ ions were calculated using the multi-configuration Dirac-Hartree-Fock (MCDHF) method and the Model-QED method. The contributions of the electron correlation effects and quantum electrodynamics (QED) effects were discussed in detail. The transition energies are in excellent agreement with the experimental results, with differences of approximately 5 cm$^{-1}$. The presently calculated $g$ factor of 2.0024905(16) for the $6S_{1/2}$ agrees very well with the available experimental and theoretical results, with a difference at a level of 10$^{-6}$. For the $5D_{3/2, 5/2}$ states, the present results of 0.7993961(126) and 1.2003942(190) agree with the experimental results of 0.7993278(3) [\textcolor{blue}{Phys. Rev. A 54, 1199(1996)}] and 1.20036739(14) [\textcolor{blue}{Phys. Rev. Lett. 124, 193001 (2020)}] very well, with differences at the level of 10$^{-5}$. 
\end{abstract}

\maketitle

\section{INTRODUCTION}

The Land$\acute{e}$ $g$ factors are very important in high-precision measurement physics, including applications in atomic clock~\cite{han2019-100,ma2024-110},
quantum computation~\cite{dietrich2010-81,hanson2008-453}, and quantum information~\cite{hucul2017-119,inlek2017-118,leibfried2004-304}. They also play a significant role in exploring high-order QED~\cite{yerokhin2017-95} and electron correlation effects~\cite{yerokhin2017electron,zinenko2023-107}, as well as in testing the accuracy of theoretical methods. For few-electron systems, such as H-like~\cite{haffner2000-85,hannen2019-52,morgner2023-622}, Li-like~\cite{glazov2019-123,glazov2005-235,kosheleva2022-128} systems, both theoretical and experimental results have achieved remarkable agreement~\cite{haffner2000-85,verdu2004-92,sturm2011-107,sturm2013-87,wagner2013-110}. The two-loop QED effects~\cite{glazov2005-235,cakir2020-101,jentschura2009-79,morgner2023-622}, screened QED effects~\cite{glazov2004-70}, higher-order electron correlation effects~\cite{sturm2011-107,aleksandrov2018-98}, and higher-order nuclear recoil effects~\cite{sailer2022-606,glazov2020-101,malyshev2019-100} have been identified as crucial. However, for many-electron systems, high-precision theoretical calculations remain challenging due to the complexity of electron correlations. 

Ba$^+$ is a promising candidate for quantum computing, quantum information, and atomic clock research~\cite{sherman2005,arnold2020-124}. Its ground state is 6$s$, and the metastable excited $5d$ states have lifetimes of 48 and 32 s for the 5$D_{3/2}$ and 5$D_{5/2}$ states, respectively~\cite{gurell2007-75,nagourney1986-56}. The $6S_{1/2} - 5D_{5/2}$ transition is a clock transition with a high quality factor of 5.4 $\times$ 10$^{15}$. The $g$ factor of the ground state 6$s_{1/2}$ has been measured with an uncertainties at the level of $10^{-7}$~\cite{lindroth1993-47,marx1998-4,knab1993-25}. Recently, Arnold \textit{et al.}~\cite{arnold2020-124} and Hoffman \textit{et al.}~\cite{hoffman2013-88} have measured the $g$ factor of the $5D_{5/2}$ state in Paul traps, achieving uncertainties at the levels of $10^{-7}$ and $10^{-6}$, respectively.

Theoretically, Lindroth \textit{et al.} have calculated the $g$ factor of the ground state $6S_{1/2}$ of Ba$^{+}$~\cite{lindroth1993-47}, and their theoretical results are in agreement with the most accurate experimental results at the level of $10^{-6}$. However, for the excited state of $5D_{3/2}$, there is only one theoretical result available from Ref.~\cite{knoll1996-54}, which agrees with the most accurate experimental results at a level of $10^{-4}$. To our knowledge, there are no theoretical results for the $5D_{5/2}$ state available for comparison with recent experimental data. Therefore, highly accurate theoretical calculations of the $g$ factors for these excited states are required.

In this work, the wave functions, energy levels, and $g$ factors of the $6S_{1/2}$ and $5D_{3/2,5/2}$ states of Ba$^{+}$ were calculated using the multi-configuration Dirac-Hartree-Fock (MCDHF) method and the Model-QED method. The electron correlation effects, Breit interactions, and QED effects were discussed in detail.

\section{Theoretical methods}

The MCDHF method is described in detail in Refs.~\cite{fischer2016-49,jonsson2017-5,grant2007-173}, here, we only provide a brief overview.
%
In this method, the atomic state wave function (ASF) for a given parity $P$, total angular momentum $J$ and its projection on the Z-axis $M$ is expressed as a linear combination of symmetry-adapted configuration state wave functions (CSFs)~\cite{jonsson2017-5}, i.e.,
\begin{eqnarray}\label{eq3}
\Psi(\gamma P J M_{J}) =\sum\limits_{j=1}^{N_{CSFs}}c_{j}\Phi(\gamma_j P J M_{J}),
\end{eqnarray}
where $c_j$ is the mixing coefficient and $\gamma$ represents other quantum numbers of the corresponding states. The single electron orbitals are optimized through the relativistic self-consistent method. Subsequently, the mixing coefficients are obtained by performing relativistic configuration interaction (RCI) calculations, i.e., by diagonalising the Hamiltonian matrix of the configuration state space. In this step, the Breit interaction, QED effects, and nuclear recoil corrections can be added in the Hamiltonian matrix. 

The Breit interaction is written as~\cite{fischer2016-49,jonsson2017-5,rodrigues2000-63},
\begin{eqnarray}
	H_{\rm Breit}&&=-\sum_{i<j}^{N}[\bm{\alpha}_{i}\cdot\bm{\alpha}_{j}\frac{cos(\omega_{ij}r_{ij}/c)}{r_{ij}}\nonumber\\
	&&+(\bm{\alpha}_{i}\cdot{\nabla}_{i})(\bm{\alpha}_{j}\cdot{\nabla}_{j})\frac{cos(\omega_{ij}r_{ij}/c)-1}{\omega^2_{ij}r_{ij}/c^2}],
\end{eqnarray}
where $\boldsymbol{\alpha}_{i}$ is the Dirac matrix consisting of three 4$\times$4  matrices,
$\nabla$ is the momentum operator, $c$ is the speed of light, and $r_{ij}$ is the distance between electron $i$ and electron $j$. The frequency $\omega_{ij} $ corresponds to the virtual photon exchanged between the electrons. The value of $\omega_{ij} $ can be the difference of single electron energies. If $\omega_{ij} \to 0$, the operator reduces to the frequency-independent Breit interaction.


The contribution of QED were calculated using the Model-QED method~\cite{shabaev2015-189,shabaev2013-88,shabaev2018-223}. This method has been successfully applied to studies of various atomic systems~\cite{malyshev2022-106,anisimova2022-106,tupitsyn2021-129,shabaev2020-101,volotka2019-100,si2018-98,machado2018-97,yerokhin2017-96,tupitsyn2017-408,tupitsyn2016-117,BingbingLi-2024}.
The QED effect is represented by the sum of the self-energy and the vacuum polarization~\cite{shabaev2015-189,shabaev2013-88,shabaev2018-223}. The vacuum polarization was calculated by the sum of the Uehling potential~\cite{fullerton1976-13} and the Wichmann-Kroll potential~\cite{fainshtein1991-24}. The self-energy operator is written as~\cite{shabaev2015-189,shabaev2018-223}
\begin{equation}\label{SE-model}
{h^{{\rm{SE}}}} = h_{{\rm{loc}}}^{{\rm{SE}}} + h_{{\rm{nloc}}}^{{\rm{SE}}}.
\end{equation}
Here, $ h^{SE}_{loc}$ is the quasi-local operator, while $h^{SE}_{nloc}$ denotes the nonlocal operator, these two operators are depicted in detail in Refs.~\cite{shabaev2015-189,shabaev2018-223}.

The $g$ factor can be written as~\cite{cheng1985-31}
\begin{eqnarray}
\label{g1}
g=\frac{1}{2 \mu_B} \frac{\left\langle\Psi(\gamma P J)\left\| \textbf{\emph{N}} ^{(1)}\right\| \Psi(\gamma P J)\right\rangle}{\sqrt{J(J+1)(2 J+1)}},
\end{eqnarray}
where the operator
\begin{eqnarray}
\label{nopre}
\textbf{\emph{N}} ^{(1)}=-i\sum_j \sum_{q=0, \pm 1}  \sqrt{\frac{8 \pi}{3}} r_j \alpha _j \cdot \textbf{\emph{Y}} _{1 q}^{(0)}\left(\hat{\textbf{\emph{r}} }_j\right).
\end{eqnarray}
Here, $\textbf{\emph{Y}} _{1 q}^{(0)}$ is the vector spherical harmonic~\cite{akhiezer1965}.
The QED correction due to the anomalous magnetic moment of the electron can be written as~\cite{cheng1985-31,akhiezer1965},
\begin{eqnarray}\label{g2}
\Delta g_{Q E D}=\frac{(g_s-2)\left\langle\Psi(\gamma P J)\left\|\sum_j \Delta \textbf{\emph{N}}_j ^{(1)}\right\| \Psi(\gamma P J)\right\rangle}{2{\sqrt{J(J+1)(2J+1)}}},
\end{eqnarray}
where $g_{s}$= 2.0023193~\cite{shabaev2002-65}. The operator $\Delta \textbf{\emph{N}}_j^{(1)}=\beta_j\mathbf{\Sigma}_{j}$ and 
$\mathbf{\Sigma}_{j} $ is the relativistic spin matrix.

\section{results and discussion}

\subsection{energy levels}

In order to accurately deal with the effects of electron correlation, three different correlation models, designated as Model A, Model B, and Model C, were employed. In Model A, the calculations were performed using the single-reference configurations ${5s^25p^{6}6s}$ and ${5s^25p^{6}5d}$, with the $5p$, $5d$, and $6s$ electrons set as active electrons. The configuration space was subsequently generated through single and double (SD) excitations from the occupied orbitals of the reference configurations to unoccupied (virtual) orbitals. The virtual orbitals were constrained to principal quantum numbers $n_{max} \leq 9$ and orbital quantum numbers $l$ $\leq$ 5 (i.e., the angular symmetries are $s, p, d, f, g$). In this model, the core-valence (CV) electron correlation effects between the $5p$ core electrons and the $6s$ or $5d$ valence electrons were considered. Model B was based on Model A, and included the $5s$ core electrons as active electrons. This model extended Model A by incorporating the CV electron correlations between the $5s$ core electrons and the $6s$ and $6d$ valence electrons, as well as the core-core (CC) electron correlations between the $5s$ and $5p$ electrons. The third model, Model C, incorporated higher-order electron correlations through SD excitations from multi-reference (MR) configurations. The configurations that significantly contributed to the wave function were selected as reference configurations. These reference configurations included $5s^{2}_{2}5p^{6}_{2}6s_{1}$, $5s^{2}_{2}5p^{6}_{2}5d_{1}$, $5s^{2}_{1}5p^{4}_{2}5d_{1}6s_{1}6d_{1}$, $5s^{2}5p^{5}_{2}5d_{1}6p_{1}$,  $5s^{2}5p^{4}_{1}5d^{2}_{2}6s_{1}$  and $4d^{9}5s^{2}5p^{6}6d_{1}6s_{1}$. The subscript denotes the maximum number of electrons that can be excited. Table~{1} lists the numbers of CSFs that expand with the principal quantum number $n$. It should be noted that in Model C, the configurations that do not interact with $6S_{1/2}$ and $5D_{3/2,5/2}$ were eliminated by using the program "$rcsfinteract$" in the GRASP2018~\cite{GRASP2018}. In this model, the $4d$ orbital was open, allowing the calculation to include the CC and CV correlations of the $4d$ electrons. 

Table~{2} shows the convergence of the transition energies with the expansion of the configuration space. It can be seen that the transition energies calculated in each model converged very well when $n_{max}$ was increased to 9.
To illustrate the convergence of the different correlation models, Fig.~1 shows the difference of the transition energies between the calculated results for each correlation model with $n_{max} = 9$ and the NIST standard values that are 4874, 5675, and 801 cm$^{-1}$ for the $5D_{3/2}-6S_{1/2}$, $5D_{5/2}-6S_{1/2}$, and $5D_{5/2}-5D_{3/2}$ transitions, respectively. It is evident that Model C provides very good results. The difference between Dirac-Fock (DF) and Model A indicates the CV and CP correlations for the 5$p$ orbitals, which accounted for $-10$\% of the transition energy. The contribution of CV and CC correlations involving the 5p orbital is about $-6$\%, which was obtain by the difference between Model A and Model B. Similarly, the contribution of CV and CC correlations involving the 4d orbital, along with partial triple- and quadruple-excitations, is $-2$\% which was derived from the difference between Model B and Model C.

\begin{table}
\begin{small}

\caption{\label{tab-number} The number of CSF for the $6S_{1/2}$, $5D_{3/2,5/2}$ 
states  of Ba$^{+}$ ions in Model A, B, and C.}
\centering
\setlength{\tabcolsep}{4.5mm}{}
\begin{tabular}{cccc}
\hline

 $n_{max}$     &        $6S_{1/2}$     &          $5D_{3/2}$ &$5D_{5/2}$         \\
 \cline{1-4}

  &         & Model A &         \\
6 & 3 409    & 6 160    & 7 820    \\
7 & 7 976    & 14 397   & 18 219   \\
8 & 14 525   & 26 188   & 33 058   \\
9 & 23 056   & 41 533   & 52 337   \\
 \cline{1-4}
  &         & Model B &         \\
6 & 6 331    & 11 447   & 14 550   \\
7 & 14 889   & 26 877   & 34 016   \\
8 & 27 125   & 48 893   & 61 690   \\
9 & 43 039   & 77 495   & 97 572   \\
 \cline{1-4}
  &         & Model C &         \\
6 & 194 770  & 370 667 & 482 445  \\
7 & 442 030  & 876 772 &1 170 127 \\
8 & 780 002  & 1 591 337 & 2 153 107\\
9 & 1 208 686 & 2 514 362& 3 431 385 \\
\hline
\end{tabular}
\end{small}
\end{table}

\begin{table}
\begin{small}

\caption{\label{tab-comver} The convergence of transition energies  of the $6S_{1/2}$, $5D_{3/2,5/2}$ states of Ba$^{+}$ ions calculated using Model A, B,and C, respectively. (in cm$^{-1}$)}
\centering
\setlength{\tabcolsep}{1.5mm}{}
\begin{tabular}{cccc}
\hline
 $n_{max}$  & 5$D_{3/2}$ $-$ 6$S_{1/2}$    &  5$D_{5/2}$ $-$ 6$S_{1/2}$   &  5$D_{5/2}$ $-$ 5$D_{3/2}$  \\
\hline
                         &                &        Model A       &             \\
6                        & 4887           & 5659          & 771         \\
7                        & 5316           & 6081          & 766         \\
8                        & 5360           & 6124          & 764         \\
9                        & 5367           & 6131          & 763         \\
\hline
                         &                &        Model B       &             \\
6                        & 4528           & 5308          & 780         \\
7                        & 4970           & 5743          & 773         \\
8                        & 5036           & 5807          & 770         \\
9                        & 5048           & 5818          & 770         \\
\hline
                         &                &   Model C            &             \\
6                        & 4683           & 5497          & 813         \\
7                        & 4945           & 5758          & 813         \\
8                        & 4945           & 5762          & 817         \\
9                        & 4925           & 5742          & 818         \\
\hline
\end{tabular}
\end{small}
\end{table}

\begin{figure}[tbh]	
	\centering{
		\includegraphics[width=9cm, height=7cm]{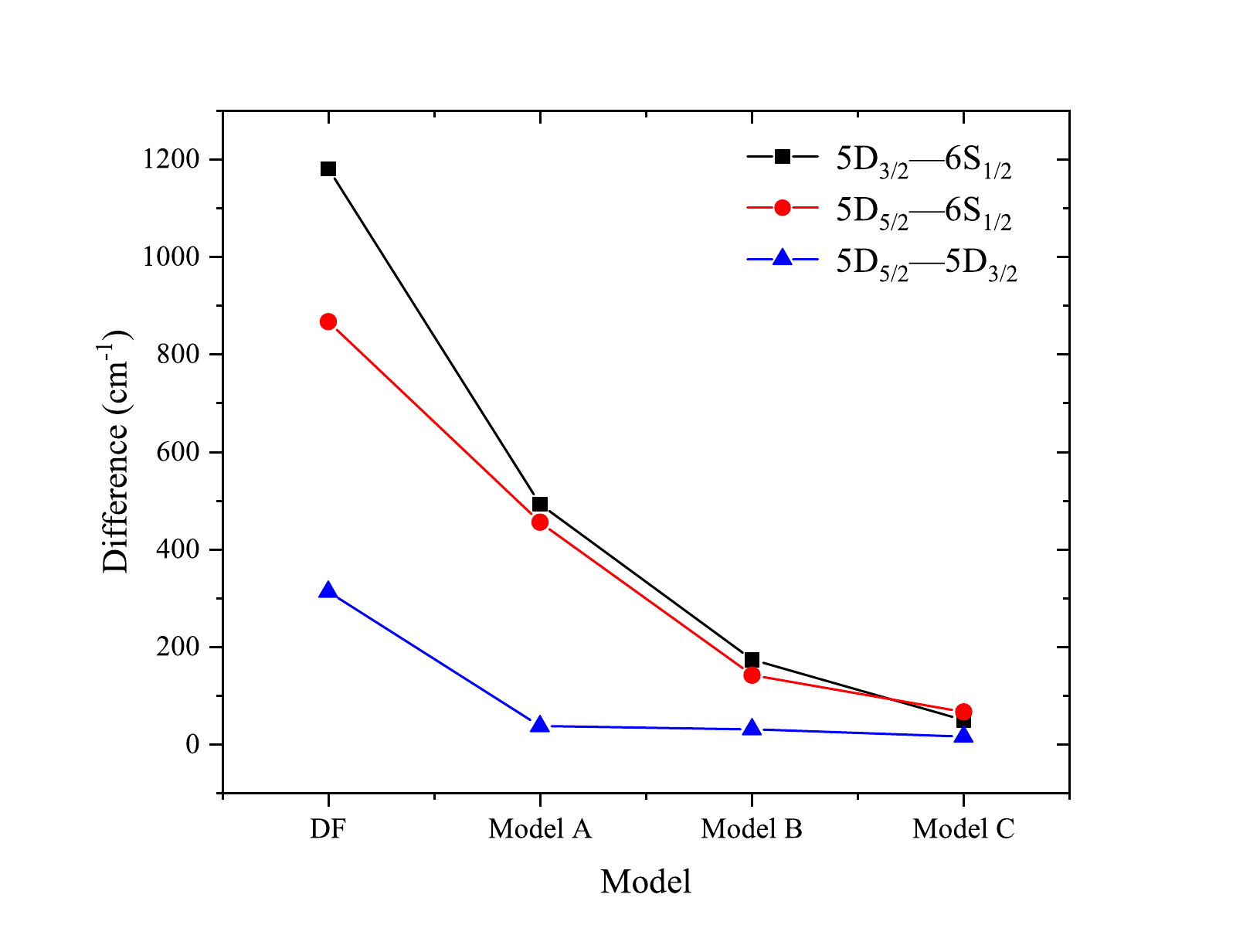} }
	\caption{\label{conver.eps}  Difference of the transition energies between the calculated results for each correlation model with $n_{max} = 9$ and the NIST standard values that are 4874, 5675, and 801 cm$^{-1}$ for the $5D_{3/2}-6S_{1/2}$, $5D_{5/2}-6S_{1/2}$, and $5D_{5/2}-5D_{3/2}$ transitions. DF is the results calculated using Dirac-Fock energies.}
\end{figure}

Then, we further calculated the Breit interactions, QED effects, and nuclear recoil by re-diagonalizing the Hamiltonian matrix on the basis of Model C with $n_{max}$=9. The matrix elements of self-energy and vacuum polarization were calculated using Model-QED program\cite{shabaev2015-189,shabaev2013-88}. These matrix elements were then added to the Hamiltonian matrix, and the QED contributions to the energy levels and wave functions were obtained through diagonalization. Table~{3} presents the individual contributions of the Breit interaction, QED effects and total transition energies for the $5D_{3/2}$ $-$ $6S_{1/2}$ and $5D_{5/2}$ $-$ $6S_{1/2}$ states, along with a comparison to some available results. Two types of Breit interactions, namely frequency-independent ($\omega \longrightarrow 0$) and frequency-dependent Breit interactions, were calculated in the present work. For the frequency-dependent Breit interactions calculations, the frequencies were still set to zero for virtual orbitals. It can be seen that the contribution of the Breit interaction comes mainly from the frequency-independent term, while the correction for $\omega\neq 0$ is negligible. Additionally, the contributions of nuclear recoil was also assessed and found to be negligible, with a magnitude of less than 0.1~cm$^{-1}$. The final energy levels labeled "Total", which included the contributions from the electronic correlation effects, Breit interactions, QED effects, show excellent agreement with the results of the NIST tabulation~\cite{NIST}.


\begin{table*}[h!t]
\renewcommand{\thetable}{3} 
\centering
\caption{Transition energy (in cm$^{-1}$) of the $6D_{1/2}$, $5D_{3/2,5/2}$ states  of Ba$^{+}$. Breit($\omega = 0$) and Breit($\omega\neq 0$) represent the contributions from the frequency-independent and  frequency-dependent interactions, respectively. The "Total" is the sum of CI, Breit($\omega\neq 0$) and QED contributions.   The "SD" represents the single-double (SD) all-order method. The "CP+CI" and "SDCC+CI"  represents orrelation-potential (CP) method and linearized singledouble-coupled-cluster (SDCC) method combine configuration-interaction (CI) techniques,respectively.}
\label{tab3}
\vskip 2mm \tabcolsep 15.5pt
\begin{tabular}{cccccc}\hline
     Contribution& $5D_{3/2}-6S_{1/2}$&$5D_{5/2}-6S_{1/2}$  &$5D_{5/2}-5D_{3/2}$    \\
      \hline
CI         & 4924.7    & 5742.3   & 817.6 \\
Breit($\omega$=0)       & $-$55.4     & $-$80.8 & $-$25.5    \\
     Breit($\omega\neq 0)$      & $-$55.6     & $-$81.6    & $-$25.9 \\
CI+Breit($\omega\neq 0)$ & 4869.0    & 5660.7& 791.7    \\

QED & 9.4       & 9.3    & $-$ 0.1  \\
Total      & 4878.4(245)    & 5670.0(248)  & 791.6(252)  \\
SD~\cite{safronova2010-81}           & 4762        & 5521        & 759\\
SDCC+CI~\cite{dzuba2014-90}           & 4869      & 5736   & 867   \\
CP+CI~\cite{dzuba2014-90}           & 4750      & 5611      & 861 \\
Exp.~\cite{NIST}       & 4873.852  & 5674.807 & 800.955 \\
\hline

\end{tabular}
\end{table*}

\subsection{Land$\acute{e}$ $g$ factor}

Ignoring the electron correlation effects, QED effects and nuclear recoil, the $g$ factor is reduced to the analytical Dirac value~\cite{zapryagaev1979-47} 
\begin{eqnarray}
g_{D} = \frac{\kappa}{2J(J+1)}(2\kappa\varepsilon_{n\kappa}-1),
\end{eqnarray}
where $\varepsilon_{n\kappa}$ is the Dirac energy of the orbital and $\kappa$ is the relativistic quantum number. For the states $6S_{1/2}$, $5D_{3/2}$, and $5D_{5/2}$ of the Ba$^{+}$, the $g_{D}$  are 1.9968281, 0.7963948, and 1.1965435, respectively.


\begin{table*}[h!t]
\renewcommand{\thetable}{4} 
\centering
\caption{ The convergence of positive-energy states to the $g$ factor of the 6S$_{1/2}$ ,5D$_{3/2}$, 5D$_{5/2}$ states for Ba$^{+}$ ions calculated using Model C. $\Delta g_{PS}$ represents the contribution of positive-energy states. $g_{D}$ is the Dirac value. }
\label{tab3}
\vskip 2mm 
\setlength{\tabcolsep}{3.0mm}{}
\begin{tabular}{ccccccccc}\hline

  \multicolumn{1}{c}{\multirow{3}*{$n_{max}$}} & \multicolumn{2}{c}{6S$_{1/2}$}& & \multicolumn{2}{c}{5$D_{3/2}$}& &\multicolumn{2}{c}{5$D_{5/2}$ }\\
\cline{2-3}
\cline{5-6}
\cline{8-9}
\multicolumn{1}{c}{\multirow{1}*{$n_{max}$}} & \multicolumn{2}{c}{$g_{D}$=1.9968281}& & \multicolumn{2}{c}{ $g_{D}$=0.7963948}& &\multicolumn{2}{c}{$g_{D}$=1.1965435 }
\\
\cline{2-3}
\cline{5-6}
\cline{8-9}
& $g$   &$\Delta$$g_{PS}$$=g$-$g_{D}$ & &$g$   &$\Delta$$g_{PS}$$=g$-$g_{D}$ &  &$g$   &$\Delta$$g_{PS}$$=g$-$g_{D}$  \\
\cline{1-9}
6 & 2.0001674 & 0.0033393 &  & 0.7998653 & 0.0034705 &  & 1.1999431 & 0.0033996 \\
7 & 2.0001617 & 0.0033336 &  & 0.7998623 & 0.0034674 &  & 1.1999382 & 0.0033948 \\
8 & 2.0001680 & 0.0033399 &  & 0.7998610 & 0.0034662 &  & 1.1999364 & 0.0033930 \\
9 & 2.0001678 & 0.0033397 &  & 0.7998604 & 0.0034656 &  & 1.1999355 & 0.0033921 \\
\cline{2-3}
\cline{5-6}
\cline{8-9}
\hline
\end{tabular}
\end{table*}

Table~{4} shows the convergence of our calculated $g$ factors using Eq.~(\ref{g1}) as the maximum principal quantum number $n_{max}$ increased in Model C. The contributions of electron correlation effects were obtained from the difference between the calculated $g$ factor and the $g_{D}$ values. When the configuration space reached $n_{max} = 9$, this contribution converges to a level of 10$^{-6}$ .

We also evaluated the contributions of the negative-energy states. The estimation method is similar to our previous calculations of the $g$ factor for Ar$^{13+}$ and Li-like Sn$^{47+}$ and Bi$^{80+}$ ions~\cite{wulei2022, liuming2024}. The negative-energy orbitals were calculated by the relativistic configuration interaction plus core polarization (RCICP) method~\cite{jiang2016-94}. We then performed the RCI calculations again, substituting the positive energy orbitals for the negative-energy orbitals. For the $6S_{1/2}$, $5D_{3/2}$, and $5D_{5/2}$ states, the contributions converged to 0.0000063, $-$0.0000009, and $-$0.0000047, respectively.

\vskip 2mm

\medskip


\begin{table*}[h!t]
\renewcommand{\thetable}{5} 
\centering
\caption{ The $g$ factor of the states 6S$_{1/2}$ ,5D$_{3/2}$, 5D$_{5/2}$ for  Ba$^{+}$ ions. The ''Total" represent the sum of $g_{D}$ , $\Delta$ $g_{PS}$, $\Delta$ $g_{NS}$ and  $\Delta$ $g_{QED}$. $\Delta$$g_{NS}$ represents the contribution of the negative-energy states. The results of $\Delta$ $g_{QED}$  are calculated using Eq.~(\ref{g2}). }
\label{tab5}
\vskip 2mm \tabcolsep 15.5pt
\setlength{\tabcolsep}{5.0mm}{}
\begin{tabular}{llll}\hline

 & 6S$_{1/2}$    & 5D$_{3/2}$    &  5D$_{5/2}$   \\
\hline
$g_{D}$   & 1.9968281     & 0.7963948    & 1.1965435             \\
$\Delta$ $g_{PS}$ & 0.0033397     & 0.0034656    & 0.0033921             \\
$\Delta$ $g_{NS}$ & 0.0000063     & $-$0.0000009   & $-$0.0000047            \\
$\Delta$ $g_{QED}$   & 0.0023205     & $-$0.0004641   & 0.0004641             \\
   Total & 2.0024905(16)     & 0.7993961(126)    & 1.2003942(190)             \\
Exp. & 2.0024922(10)~\cite{knoll1996-54} & 0.800(9)~\cite{lindroth1993-47}     & 1.20036739(24)~\cite{arnold2020-124}        \\
    & 2.00249192(3)~\cite{marx1998-4} & 0.7993278(3)~\cite{knoll1996-54} & 1.200372(4)stat(7)sys~\cite{hoffman2013-88} \\
    & 2.0024906(12)~\cite{knab1993-25} & 0.799311(4)~\cite{werth1995-59}  & 1.2020(5)~\cite{kurz2010-82}                 \\
 Theory   & 2.0024911(30)~\cite{lindroth1993-47} & 0.79946~\cite{knoll1996-54}    &                       \\

\hline
\end{tabular}
\end{table*}

\vskip 2mm

\medskip

Table~{5} lists the individual contributions to the $g$ factors from different effects and compares them with existing theoretical and experimental results. The contributions of QED effects, denoted as "$\Delta g_{QED}$", were calculated using Eq.~(\ref{g2}). For the $6S_{1/2}$ ground state, the accuracy of the total $g$ factors for all theoretical and experimental results reaches $10^{-6}$, demonstrating excellent consistency within their respective uncertainty ranges. Our calculated results show a difference of $6\times10^{-7}$ from the theoretical result reported by Lindroth \textit{et al.}~\cite{lindroth1993-47}. Notably, our results are in excellent agreement with the experimental result from Ref.~\cite{knab1993-25}, with a difference of less than $10^{-7}$.

For the excited state $5D_{3/2}$, three experimental results~\cite{lindroth1993-47,knoll1996-54,werth1995-59} and one theoretical result~\cite{knoll1996-54} are available. The difference between the two most accurate experiments~\cite{knoll1996-54,werth1995-59} is $1.6\times10^{-5}$. The discrepancy between our result and the result of the most accurate experiment~\cite{knoll1996-54} is $6.6\times10^{-5}$, representing approximately an order of magnitude improvement in the accuracy of the result reported in Ref.~\cite{knoll1996-54}. For the $5D_{5/2}$ state, three sets of experimental results are available for comparison. The difference between the two most accurate experiments~\cite{arnold2020-124,hoffman2013-88} is $5\times10^{-6}$. The difference between our theoretical results and the most recent experimental results~\cite{arnold2020-124} is $2.7\times10^{-5}$. The primary sources of this discrepancy are the higher-order electron correlation effects and higher-order QED effects.

\section{Conclusions}

 The wave function, energy levels, and $g$ factors for the $6S_{1/2}$, and $5D_{3/2,5/2}$ states of Ba$^{+}$ ions were calculated using the MCDHF method and Model-QED method. 
The electron correlation effects were investigated in detail using three different correlation models. The contribution of CV and CC correlation effect of $5p$, $5s$, and $4d$ electrons (along with partial triple- and quadruple-excitations) to the transition energies, reach $-$10\%, $-$6\%, and $-$2\%, respectively. By including Breit interactions and QED effects, the differences between the currently calculated transition energies and the NIST standard values are reduced to approximately 5 cm$^{-1}$.
In the calculation of the Land$\acute{e}$ $g$ factors, the contributions from both positive and negative energy states and QED effects were analyzed. The present results show in good agreement with the experimental results, with differences at the level of 10$^{-6}$ and 10$^{-5}$ for the $6S_{1/2}$ and $5D_{3/2,5/2}$ states, respectively.

The current calculations demonstrates significantly improved accuracy compared to existing theoretical calculations and showed good agreement with the most precise experimental results. However, discrepancies persist between theoretical and experimental values for the excited states $5D_{3/2}$ and $5D_{5/2}$, where the theoretical results lie outside the experimental error bars. The difference, on the order of $10^{-5}$, is two orders of magnitude larger than the experimental precision of $10^{-7}$. Therefore, further advanced theoretical works are still needed, particularly regarding the higher-order electron correlation effects, two-loop QED effects, and relativistic nuclear recoil effects. These effects have been proven critical in few-electron systems but remain challenging to address in many-electron systems due to computational complexity.

\section{Acknowledgments}
This work has been supported by the National key Research and Development Program of China under Grant No.2022YFA1602500, the National Natural Science Foundation of China under Grant No.12174316 ,1236040286,12404306.

\bibliography{Li}

\end{document}